%% file: root.tex
\documentclass[letterpaper, 10 pt, conference]{ieeeconf}  

\IEEEoverridecommandlockouts                              

\usepackage{amsmath}
\usepackage{cases}
\usepackage{amssymb}  
\usepackage{amsthm}
\usepackage{algorithm,algpseudocode}
\usepackage{subfigure}
\usepackage{graphicx}
\usepackage[utf8]{inputenc}
\usepackage{float}
\usepackage[open,openlevel=1]{bookmark}
\usepackage{multicol}
\usepackage{multirow}
\usepackage{booktabs}
\usepackage{siunitx}

\makeatletter
\newcommand{\SplitState}[1]{%
  \State
  \parbox[t]{\dimexpr\linewidth-\ALG@thistlm}{%
    #1\par\xdef\Split@prevdepth{\the\prevdepth}%
  }\par
  \prevdepth\Split@prevdepth
}
\makeatother

\input{misc/mathdef}

\title{\LARGE \bf
An Efficient Solution to the 2D Visibility Problem in Cartesian Grid Maps and its Application in Heuristic Path Planning
}

\author {
Ibrahim Ibrahim$^{\dagger}$,
Joris Gillis$^{\dagger}$,
Wilm Decré$^{\dagger}$,
Jan Swevers$^{\dagger}$ 
\thanks{This work has been carried out within the framework of Flanders Make's SBO project ARENA (Agile \& Reliable Navigation). Flanders Make is the Flemish strategic research centre for the manufacturing industry.}%
\thanks{$^\dagger$ Motion Estimation Control and Optimization Lab, KU Leuven.
}%
}%
        
\begin{document}

\maketitle
\thispagestyle{empty}
\pagestyle{empty}

\begin{abstract}

This paper introduces a novel, lightweight method to solve the visibility problem for 2D grids. The proposed method evaluates the existence of lines-of-sight from a source point to all other grid cells in a single pass with no preprocessing and independently of the number and shape of obstacles. It has a compute and memory complexity of $\mathcal{O}(n)$, where $n = n_{x}\times n_{y}$ is the size of the grid, and requires at most ten arithmetic operations per grid cell. In the proposed approach, we use a linear first-order hyperbolic partial differential equation to transport the visibility quantity in all directions. In order to accomplish that, we use an entropy-satisfying upwind scheme that converges to the true visibility polygon as the step size goes to zero. This dynamic-programming approach allows the evaluation of visibility for an entire grid orders of magnitude faster than typical ray-casting algorithms.
We provide a practical application of our proposed algorithm by posing the visibility quantity as a heuristic and implementing a deterministic, local-minima-free path planner, setting apart the proposed planner from traditional methods. Lastly, we provide necessary algorithms and an open-source implementation of the proposed methods.

\end{abstract}

\section{Introduction}
\label{sec:intro}
\begin{figure}[ht!]
    \centering
    \includegraphics[width=8.5cm,keepaspectratio]{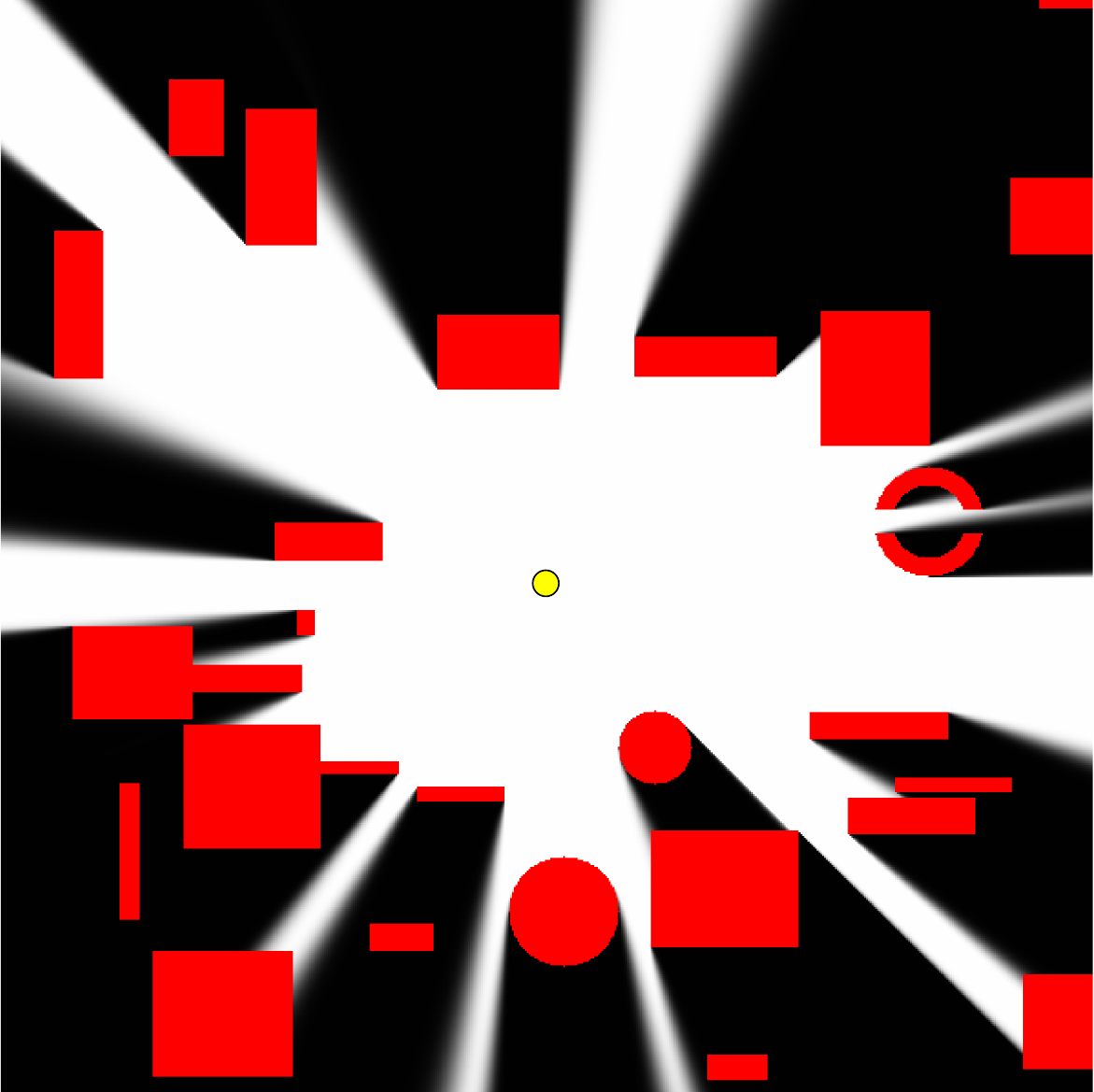}
    \caption{The 2D visibility polygon $\mathcal{U}$ computed using our algorithm. The robot position is the light source shown as a yellow dot. The white region is visible to the robot, constituting the visibility polygon. Dark regions are invisible to the robot. Obstacles are shown in red. $1000\times1000$ grid.}
    \label{fig:visibility_polygon}
\end{figure}

Visibility is crucial for robot applications as it enables the robot to make informed decisions and allows it to interact with the world in a meaningful way. Tasks such as obstacle avoidance and path planning require knowledge about the regions accessible from the robot’s position at any time instant, while other tasks such as object recognition, tracking, and manipulation necessitate maintaining a line-of-sight with the target at all times. The robot’s visibility of a point in an environment --- or the robot’s accessibility to a point --- can be defined as the existence of an uninterrupted line-of-sight between the robot and the point. The set of points that are visible to the robot at any time instant constitutes the robot’s visibility polygon. An example of a visibility polygon in a cluttered environment is illustrated in Fig.~\ref{fig:visibility_polygon}. In robotic applications, there are two predominant approaches for representing environments: grid maps and polygonal domains. In this introduction, we delve into the issue of visibility within each of these representations.

Polygonal domains constitute a simplified representation of environments where the robot, the obstacles, and the internal and external boundaries are assumed to be polygons \cite{Sack2000Domains}. This assumption, although restrictive, makes tackling the problem of 2D visibility tractable by working with polygon vertices to construct visibility polygons and/or visibility graphs. As such, 2D visibility has predominantly been explored within the context of polygonal domains, as extensively documented in the literature \cite{Davis1979ComputationalMO, Joe1987CorrectionsTL, Asano2005VisibilityOD, Heffernan1995Holes}. Nonetheless, it is essential to acknowledge the inherent limitations of these methods: they all scale with the number of obstacle vertices and edges. Furthermore, these methods often exhibit determinism and static behavior, making them less suitable for dynamic and probabilistic environments, which impose substantial computational and memory burdens. Lastly, it is noteworthy that the prevailing techniques for constructing visibility polygons and graphs cannot be readily extended to 3D environments, primarily due to the intractability associated with handling vertices in three dimensions.

Grid maps find favor among roboticists due to their inherent simplicity, user-friendly characteristics, and robustness. They provide granularity and precision, particularly in tasks demanding meticulous localization, path planning, and obstacle avoidance. They are also easy to visualize and interpret, particularly in dynamic environments. However, the realm of 2D visibility in grid map representations remains a relatively underexplored domain in the current literature. Existing methods often focus on evaluating visibility for individual pixels one at a time. However, certain applications necessitate a comprehensive understanding of visibility across the entire grid map, including autonomous path planning, obstacle avoidance, sensor placement, exploration, and area coverage. Consequently, the need arises to quantify visibility for all grid map pixels. Current methodologies rely on ray-shooting, ray casting, or voxel traversal algorithms to compute visibility grid maps, but these approaches suffer from inherent inefficiencies due to the necessity of conducting line-of-sight checks for each grid cell. Moreover, some of these methods exhibit limitations concerning scalability and complexity.

In light of these considerations, this paper addresses the understudied realm of 2D visibility in grid map representations, aiming to contribute novel solutions to this important challenge. To this end, we introduce an efficient method that computes the visibility grid map in a single pass with no assumptions, no preprocessing, and independently of the number and the shape of obstacles. It uses only raw information from a deterministic or probabilistic occupancy grid, which is either given or is constructed from sensor data such as LIDAR. We accomplish that by transporting the visibility quantity using an entropy-preserving, stable, and converging upwind scheme solution to a linear first-order hyperbolic Partial Differential Equation (PDE). The visibility quantity itself is a value ranging from 1, meaning fully visible, to 0, meaning fully invisible. This method extends to 3D and extends to curves, allowing us to quantify \emph{curves-of-sight} --- the existence of an unobstructed curve, of a generic shape, between the robot and any other point/s. The result is a \emph{curvilinear polygon} rather than a visibility polygon.

Moreover, we use the produced visibility quantity as a heuristic for path planning as an application to our visibility grid map algorithm. Even though visibility has long been used in planning methods, particularly in polygonal domains, to the best of our knowledge, it has not been directly embedded as a guiding heuristic in a Dijkstra-like algorithm, similar to the distance heuristic that guides an $A^\ast$ path planner, within the realm of grid maps.
The resulting planner builds a path by iteratively placing a new waypoint that is \emph{barely} visible to its parent waypoint while minimizing the overall path length. The heuristic can also be tuned to favor exploration of unlit/unseen regions over driving towards the target. The map is deemed completely explored once each accessible grid cell has a line-of-sight connection with at least one placed waypoint.

In this paper, we use the terms “visibility polygon” and “visibility grid map” interchangeably. The contribution of this paper is threefold:
\begin{itemize}

\item Posing visibility as a transportable quantity and transporting it using a stable and converging solution to a linear first-order hyperbolic PDE, allowing us to compute the visibility grid map and the \emph{curvilinear polygon} efficiently for single or multiple light sources. 

\item As an interesting application to our visibility grid map algorithm, we introduce the visibility heuristic in a way that drives a greedy any-angle path planner towards the target. A planner is said to be any-angle if it allows the turns in the path between two grid points to have any angle. This results in a direct point-to-point path that traverses open areas without being restricted to predefined neighbouring directions/angles at each step. The proposed planner can instead be used to explore the map, ensuring that every point in the grid is seen by (connected to) at least one waypoint.

\item Algorithms to compute the visibility polygon and to implement the heuristic path planner as well as numerical experimental results and open-source implementations \href{https://github.com/IbrahimSquared/visibility-heuristic-path-planner}{https://github.com/IbrahimSquared/visibility-heuristic-path-planner}.

\end{itemize}

\section{Related Work}
\label{sec:literature}

The visibility problem is a basic computational geometry problem \cite{Bungiu2014EfficientCO, Barbra2013Few, Chvatal1975Combinatorial}, that has been studied extensively and has multiple applications including computer graphics \cite{Newell1972Hidden}. 

\subsection{Polygonal Domains}
The computation of the visibility polygon of a point in a simple polygon as studied by Davis and Benedikt \cite{Davis1979ComputationalMO} has an $\mathcal{O}(V^{2})$  time complexity where $V$ is the number of vertices. More efficient algorithms came along afterwards which simplified it down to $\mathcal{O}(V)$ \cite{Joe1987CorrectionsTL}. For the case of a simple polygon with holes, Asano et al. \cite{ Asano2005VisibilityOD} provided a solution that has a $\mathcal{O}(V\log{}V)$ time complexity. A variant that scales with the number of obstacles $h$ also as $\mathcal{O}(V + h\log{}h)$  was presented by Heffernan et al. \cite{Heffernan1995Holes}.

The solution to answering visibility polygon queries for a generic point, called visibility polygon query problem, is approached by preprocessing the environment polygon and constructing data structures which make querying visibility polygons more efficient. For example, Bose et al. \cite{Bose2002Queries} address preprocessing in $\mathcal{O}(V^{3}\log{}V)$ time with data structures of size $\mathcal{O}(V^{3})$ to answer visibility polygon queries in $\mathcal{O}(\log{}V + k)$  where $k$ is the size of the visibility polygon. Recovering the number of vertices visible from the source in this approach requires an added $\mathcal{O}(\log{}V)$ time.

Visibility graph path planning assumes that obstacles are 2D polygons and creates the visibility graph by using knowledge on obstacle vertices and edges. It then finds the optimal path using efficient graph path planners such as $A^\ast$ or Dijkstra \cite{Lozano1979Polyhedral}. There exist  multiple ways of constructing the visibility graph of a polygonal domain. Pocciola and Vegter \cite{Pocchiola1996TopologicallySV} provide a method that constructs a visibility graph in $\mathcal{O}(V + E)$  time and $\mathcal{O}(V)$  space where $E$ is the number of obstacle edges. A method that performs triangulation of the free space of the polygonal domain first was put forth by Kapoor and Maheshwari \cite{Kapoor2000VGraph} that has a compute complexity of $\mathcal{O}(T + E + h\log{}V)$  time and $\mathcal{O}(E)$  space where $T$ is the triangulation time. Other methods relying on triangulation can handle a large number of vertices and edges dynamically \cite{LCT2014} but share similar complexity limitations. 

\subsection{Grid Map Representation}
Ray casting, ray shooting, and voxel traversal are the most widely adopted methods to quantify visibility, albeit in a binary way, for robot applications in grid environments. All said methods visit each cell more than once and therefore perform redundant computations. For example, in Next-Best-View (NBV) planning schemes, where the purpose is to plan the next camera pose in a way that maximizes the likelihood of obtaining the highest amount of information from sensors such as visibility and matching of image features, authors often fallback to using ray-casting \cite{Zeng2020View}, \cite{Wu2015Active}, \cite{McGreavy2016NBV}. In \cite{Santos2020Autonomous}, the authors perform autonomous scene exploration by encoding visibility or lack thereof in a dual OcTree structure by simulating ray-casting as part of their NBV planning approach.

The notions of visibility and obstruction are also commonly required for multi-agent hide-and-seek or pursuit-evasion applications. In \cite{Bowen2020Emergent}, the authors endow the agents with sight by using an array of 30 simulated rays arranged uniformly around the agent that cover a cone of interest, representing the agent’s field of view, similar to a simulated lidar. Tandom and Karlapalem \cite{Tandon2018MedusaTS} also trace uniformly spaced rays that emit from agents to represent the agent’s associated visibility region in simulation. Agents not within visibility cones are masked. One commonly adopted voxel traversal algorithm was presented by Amanatides et al. \cite{Amanatides1987Traversal}.

Occlusion avoidance is another common theme for utilizing the notions of visibility and obstruction. The point in such applications is for the robot to maintain a line-of-sight with the target of interest, e.g., when using manipulators for videography. In \cite{Ibrahim2022Occlusion}, the authors present a single pass method to provide a differentiable quantification of visibility in the form of a scalar field for both 2D and 3D grids, but such a method only provides an approximation of visibility which fails in tight or oblique corridors. The formulation in \cite{Ibrahim2022Occlusion} is not formal and is based on a simple geometric approach that fails in some cases. As such, it is deemed unreliable and not robust. On the other hand, Nageli et al. \cite{Nageli2017Aerial} implement a fast visibility check that is limited to evaluating a single point against an ellipsoidal approximation of obstacles. Lastly, Allaire et al. \cite{Allaire2022Accessibility} tackle the problem of visibility, or accessibility as they call it, by evaluating the Eikonal equation to every single point and comparing it to the geodesic shortest distance. They extend this to accessibility from a surface by discretizing the surface to a set of points and repeating the process for each point. In order to smoothen their binary solution, they consider obstacles as smooth Heaviside functions. Such an approach entailing a multi-step process is highly computationally demanding.

In \cite{OPM2020}, Farias and Kallmann present a method to compute optimal path maps with a $\mathcal{O}(\frac{n}{c}V^{2})$ complexity where $c$ is the number of GPU cores utilized for the computation. In order to compute the visibility polygon, they use a geometry shader to draw into a stencil buffer three triangles behind every obstacle line-segment that is front-facing with respect to every source point. This process is repeated for all grid cells $n$ for every obstacle line-segment and for every source point. As such, their proposed method scales with the number of obstacles in addition to the number of grid cells $n$. Moreover, such a procedure involves costly algebra routines and is not limited to simple arithmetic operations as our method.

Once again, this paper proposes a method to compute the visibility grid map for grid-based environment representation. The proposed visibility algorithm is formalized and is independent of $V$, $h$, and $E$. Most importantly, it scales linearly with the size of the grid $n = n_{x} \times n_{y}$. To the best of our knowledge, such a formal and efficient method that evaluates visibility for grid maps does not exist in literature.

\section{Visibility}
\label{sec:visibility}

\subsection{Definition}

Numerically, visibility is a scalar property of a point in a scene consisting of light sources and obstacles,
describing the point's degree of illumination by a certain source with a real number in the closed interval $\mathcal{V}=[0, 1]$.

\subsection{Formulation}
\label{subsec:pde}

Consider the linear first-order hyperbolic PDE of the form
\begin{equation}
    a(x,y) \frac{\partial{u(x,y)}}{\partial{y}} + b(x,y) \frac{\partial{u(x,y)}}{\partial{x}} = 0,
    \label{eq:advectionVectorField}
\end{equation} 
where $u(x,y) \in V$ is the visibility at every grid position $(x,y)$ and $a(x,y)$ and $b(x,y)$ are non-zero scalar components of the vector field describing the direction of rays starting from the source and going out in all directions. This formulation can be further simplified to
\begin{equation}
    \frac{\partial{u(x,y)}}{\partial{y}} + c(x,y) \frac{\partial{u(x,y)}}{\partial{x}} = 0,
    \label{eq:advectionScalarField}
\end{equation} 
where $c(x,y)$ is the quotient of $b(x,y)$ by $a(x,y)$ --- a scalar field describing the direction of rays. Equation \eqref{eq:advectionScalarField} is often referred to as the linear transport equation or the advection equation, which is a special case of the former \cite{Evans2010}.

\subsection{Solution}
\label{subsec:pdeSolution}
For our case, the vector field describing the directions of the rays has the normalized components
\noindent\begin{minipage}{.5\linewidth}
\begin{equation}
  a(x,y) = \frac{x}{\sqrt{x^{2} + y^{2}}},
  \label{eq:a_component}
\end{equation}
\vspace{\dp0}%
\end{minipage}%
\begin{minipage}{.5\linewidth}
\begin{equation}
  b(x,y) = \frac{y}{\sqrt{x^{2} + y^{2}}},
  \label{eq:b_component}
\end{equation}
\vspace{\dp0}%
\end{minipage}
meaning our scalar field becomes
\begin{equation}
  c(x,y) = \frac{y}{x}.
  \label{eq:c}
\end{equation}

The analytic solution for \eqref{eq:advectionScalarField} is trivial, but a numerical solution applicable to grids may in some cases be challenging. We adopt the first-order upwind scheme \cite{patankar1980numerical} which is a numerical discretization method for solving hyperbolic PDEs. Such a scheme is based on the idea of using the direction of the flow to determine the direction of the numerical approximations. It approximates the solution by taking into account only the information from the side of the flow from which the flow is coming. Applying it to \eqref{eq:advectionScalarField} yields
\begin{equation}
  \frac{u_{i}^{j+1} - u_{i}^{j}}{\Delta y} + c(x, y) \frac{u_{i}^{j} - u_{i-1}^{j}}{\Delta x} = 0,
  \label{eq:upwind}
\end{equation}
where $i$ and $j$ are the steps along $x$ and $y$ respectively and $\Delta x$ and $\Delta y$ are the step sizes in both dimensions respectively. The light source is initially at $(i,j)$ = $(0,0)$ in this case. From \eqref{eq:upwind} we write our update equation as
\begin{equation}
  u_{i}^{j+1} = u_{i}^{j}
  - C (u_{i}^{j} - u_{i-1}^{j}),
  \label{eq:update_1}
\end{equation}
where $C$ is the dimensionless $\mathbf{Courant}$ number defined as
\begin{equation}
  C = \frac{c(x, y) \Delta y}{\Delta x}.
  \label{eq:C}
\end{equation}

For our PDE to converge and maintain stability, it is \emph{necessary} to satisfy the Courant–Friedrichs–Lewy (CFL) condition \cite{Lewy1928}
\begin{equation}
  C = \frac{c(x, y) \Delta y}{\Delta x} \leq C_{max},
  \label{eq:CFL}
\end{equation}
where $C_{max}$ depends on the discretization technique. For an upwind scheme, we have stability for $C_{max} = 1$ and $c(x,y) \geq 0$ \cite{hirsch2007numerical}. It is then \emph{sufficient} to perform numerical experiments to validate stability and convergence.

For the case where we have a uniform grid with $\Delta x = \Delta y = 1$, condition \eqref{eq:CFL} will not be held whenever $y > x$ in \eqref{eq:c}. In such a case, we adapt our update step \eqref{eq:update_1} as
\begin{equation}
  u_{i+1}^{j} = u_{i}^{j}
  - \frac{1}{C} (u_{i}^{j} - u_{i}^{j-1}).
  \label{eq:update_2}
\end{equation}

%

Equations \eqref{eq:update_1} and \eqref{eq:update_2} allow us to propagate initial visibility values, but they do not account for internal obstacles/boundaries which themselves affect visibility. Therefore, after transporting visibility into a new cell, we multiply the resulting value with the complement of the occupancy value held at the cell's coordinates (e.g., if a cell $(i, j)$ is $70\%$ occupied $\implies$ it has a $30\%$ occupancy complement, or visibility). Lastly, we treat the cases where $x = 0$ and $y = 0$ as special boundary conditions, allowing us to write the complete Alg.~\ref{alg:VisibilityAlgorithm} for the first quadrant relative to the light source position. Other quadrants can be updated in the same fashion with the difference being some sign and bound limit changes. We increment the indices at a step size in each dimension, meaning $\Delta x$ and $\Delta y$ are already factored in Alg.~\ref{alg:VisibilityAlgorithm}. The initial light strength can be set as well as a decay factor $\alpha$ that diminishes the visibility exponentially.
 
\begin{algorithm}
\caption{Visibility Algorithm for the First Quadrant}
\label{alg:VisibilityAlgorithm}
\textbf{Inputs:} 
$\mathcal{O} \gets$ Occupancy grid complement \\
\hspace*{10mm} $(l_{x}, l_{y}) \gets$ Light source position \\ 
\hspace*{10mm} $(n_{x}, n_{y}) \gets$ Grid dimensions \\
\hspace*{10mm} LightStrength $\gets$ Light source strength, $V$ \\
\hspace*{10mm} $\alpha \gets$ Visibility decay factor, $V$ \\
\textbf{Output: } $ \mathcal{U} \gets$ Visibility polygon, $V^{n_x \times n_y}$
\begin{algorithmic}[1]
    \Procedure{GetVisibilityPolygon}{Inputs}
    \State $m_{x} = n_{x} - l_{x}$, $m_{y} = n_{y} - l_{y}$
    \For{$i = 0 \text{ to $m_x$ }$ and $j = 0 \text{ to $m_y$ }$}
        \State $p_{x} = l_{x} + i$, $p_{y} = l_{y} + j$
        \If{$i = 0 \land j = 0$}
            \State $v \gets$ LightStrength
        \ElsIf{$i = 0$}
            \State $v \gets \mathcal{U}_{p_{x}}^{p_{y}-1}$
        \ElsIf{$j = 0$}
            \State $v \gets \mathcal{U}_{p_{x}-1}^{p_{y}}$
        \ElsIf{$i = j$}
            \State $v \gets \mathcal{U}_{p_{x}-1}^{p_{y}-1}$
        \ElsIf{$i > j$}
            \State $c \gets \frac{p_{y} - l_{y}}{p_{x} - l_{x}}$
            \State $v \gets \mathcal{U}_{p_{x}-1}^{p_{y}} - \frac{c \Delta y}{\Delta x} (\mathcal{U}_{p_{x}-1}^{p_{y}} - \mathcal{U}_{p_{x}-1}^{p_{y}-1})$
        \ElsIf{$i < j$}
            \State $c \gets \frac{p_{x} - l_{x}}{p_{y} - l_{y}}$
            \State $v \gets \mathcal{U}_{p_{x}}^{p_{y}-1} - \frac{c \Delta x}{\Delta y} (\mathcal{U}_{p_{x}}^{p_{y}-1} - \mathcal{U}_{p_{x}-1}^{p_{y}-1})$
        \EndIf
        \State $\mathcal{U}_{p_{x}}^{p_{y}} \gets  v \times \mathcal{O}_{p_{x}}^{p_{y}} \times \alpha$ 
    \EndFor
    \State \textbf{return} $\mathcal{U}$
    \EndProcedure
\end{algorithmic}
\end{algorithm}

\subsection{Properties, Efficiency and Extension to Curves and 3D}
\label{subsec:discussion}
The resultant $\mathcal{U}$ from Alg.~\ref{alg:VisibilityAlgorithm} contains values ranging from $0$ to $1$ representing the visibility quantity. Using rough step sizes results in dispersion, which arises naturally from the upwind scheme, but such a dispersion decreases as the step sizes are refined. We can obtain the binary visibility polygon with the operation $\Tilde{\mathcal{U}} = \mathcal{U} \geq 0.5$. The threshold value $0.5$ can be set as desirable. A lower threshold means an under-estimation of the visibility polygon, whereas a higher one means an over-estimation. For the path planner application discussed in Section ~\ref{sec:planner}, a step size of 1 provides a sufficiently good approximation of the visibility polygon.

\begin{figure}[t]
    \centering
    \begin{subfigure}{}
        \includegraphics[width=4cm]{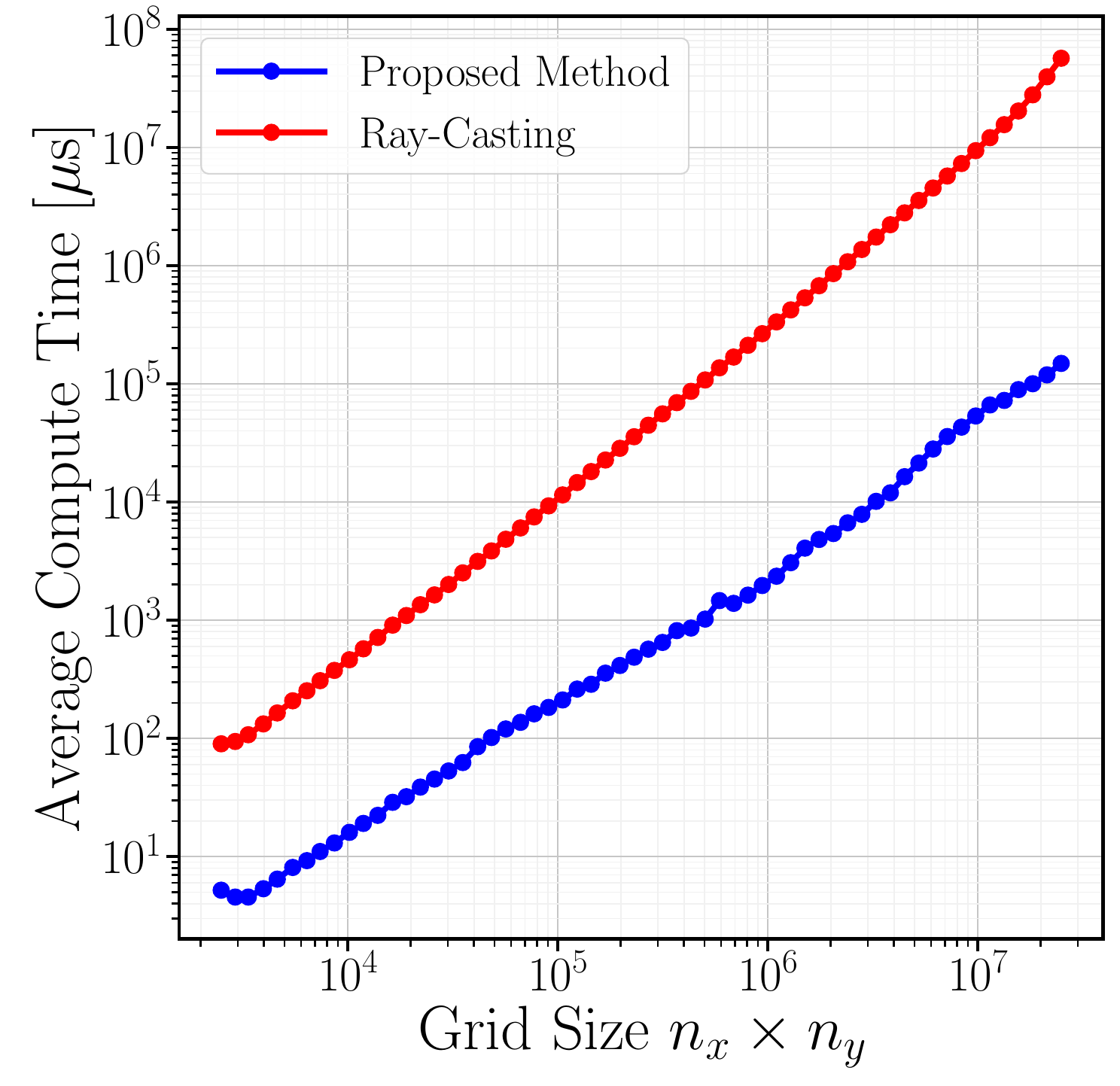}
    \end{subfigure}\hfil
    \begin{subfigure}{}
        \includegraphics[width=4cm]{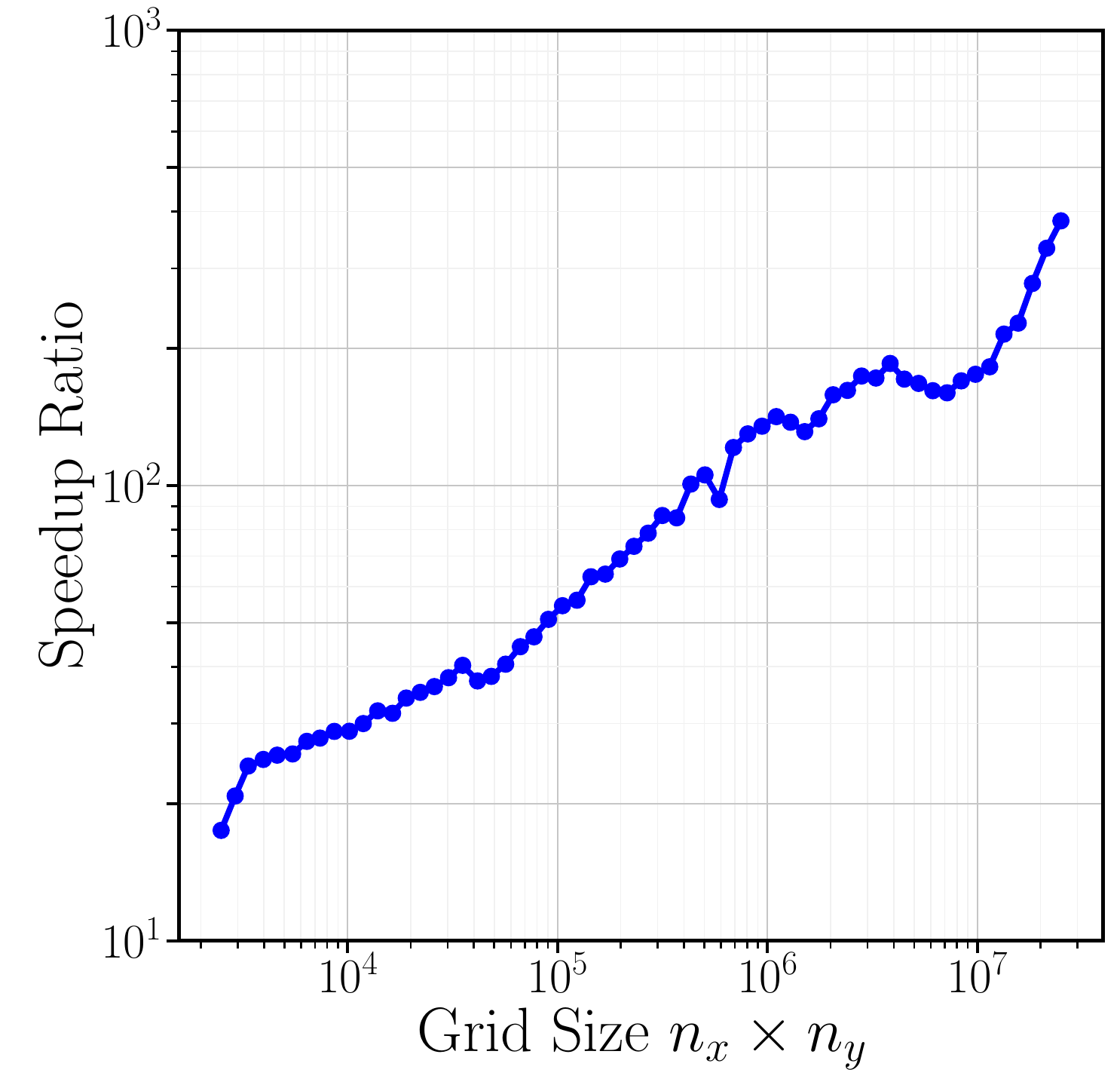}
    \end{subfigure}\hfil
    \caption{Log-log plot of average compute time of Alg.~\ref{alg:VisibilityAlgorithm} (left) with respect to the grid size $n_{x}\times{}n_{y}$ and in comparison to that of ray-casting. Error bars are insignificant/negligible over the 20 repetitions of each experiment. The speedup achieved using Alg.~\ref{alg:VisibilityAlgorithm} over ray-casting is illustrated to the right.}\label{fig:linear_complexity}
\end{figure}

Alg.~\ref{alg:VisibilityAlgorithm} requires at most 2 additions, 4 subtractions, 2 divisions, and 2 multiplications operations per grid cell and requires one array of the grid size in memory, making it extremely cheap and scalable relative to methods in literature and attractive for applications with low computational and memory capacity. A sample $100\times100$ grid can be evaluated at a constant time of around \SI{18}{\micro\second} in C++ on an Intel\textregistered~Core\texttrademark~i9--13980HX , which scales \emph{linearly} with the grid size $n_{x} \times n_{y}$, unlike similar methods in literature that additionally scale with the number of obstacle vertices in the environment. In order to demonstrate the latter, we carried out 60 sets of experiments, with each set comprising 20 repetitions, covering a range of logarithmically-sampled environment sizes. The range of sizes varied from $50 \times 50$ to $5000 \times 5000$. The resulting C++ average compute time is shown in Fig.~\ref{fig:linear_complexity} and is \textbf{independent} of the number and shape of obstacles. This single-pass dynamic-programming approach is also considerably faster and more efficient than performing typical line-of-sight checks to every single pixel in the grid map. The proposed method significantly reduces the number of required floating-point operations for evaluating visibility across the entire grid compared to adopting a voxel traversal algorithm like [21]. We illustrate the result of performing the same set of experiments using typical ray-casting in Fig.~\ref{fig:linear_complexity}. Our proposed method runs up to $100\times{}$ faster than ray-casting for an empty $1000\times{}1000$ grid, and up to $400\times{}$ faster for a $5000\times{}5000$ one. Such speedups decrease for denser environments since ray-casting terminates on collisions. In this case, one can adapt ~\eqref{alg:VisibilityAlgorithm} with termination conditions. More implementation details are available in our provided C++ open source code.

\begin{figure}[t]
    \centering
    \begin{subfigure}{}
        \includegraphics[width=4cm]{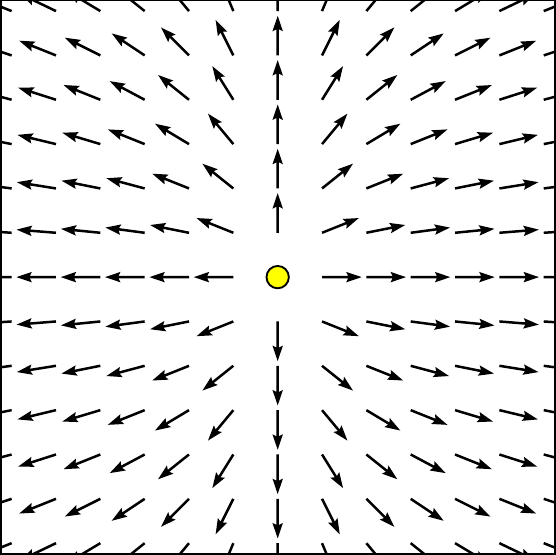}
    \end{subfigure}\hfil
    \begin{subfigure}{}
        \includegraphics[width=4cm]{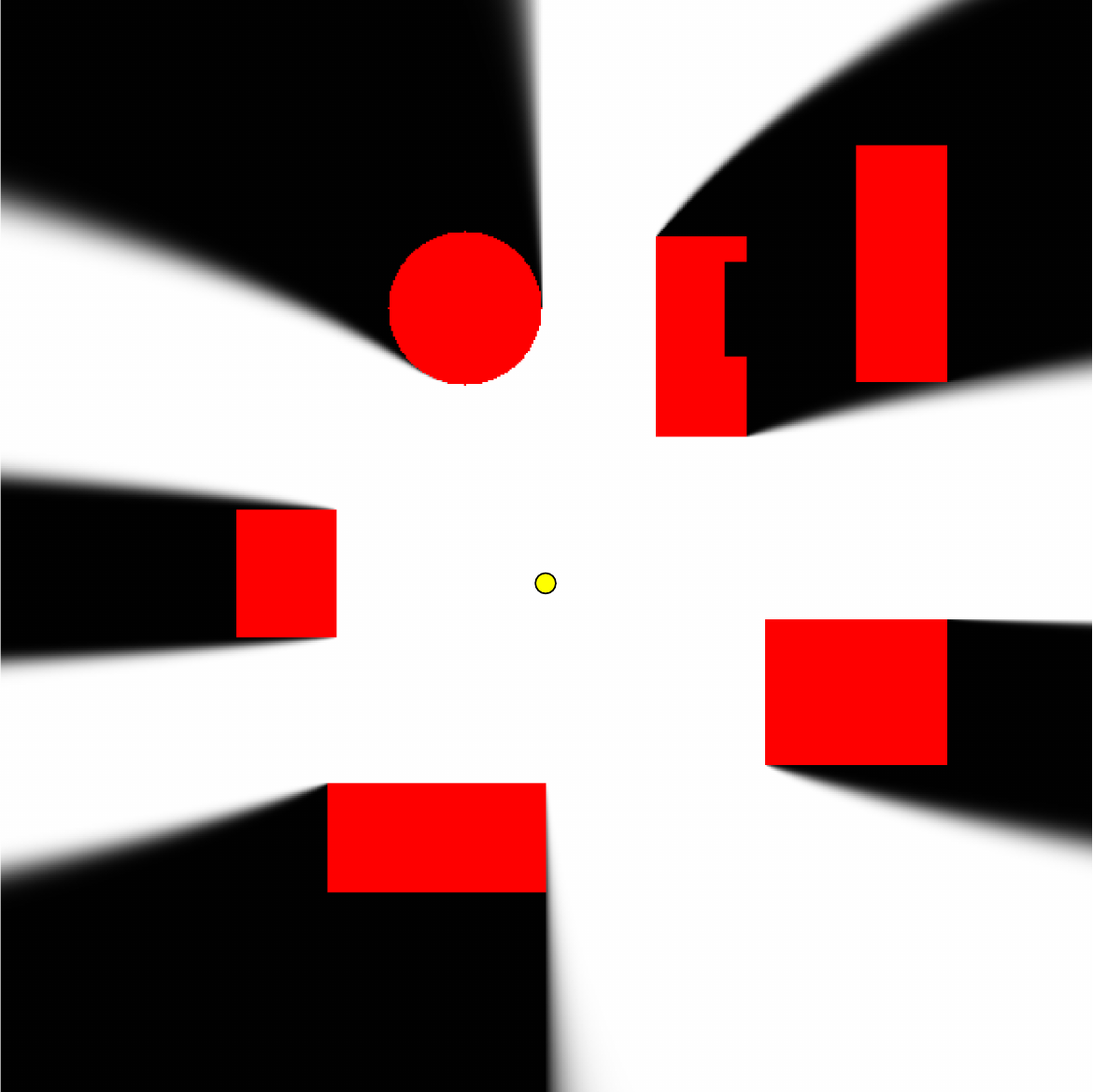}
    \end{subfigure}\hfil
    \caption{Quiver plot on left showing the vector field, centered around the light source, that governs the resulting behaviour of the \emph{curves-of-sight}. Resulting scalar field is $\frac{y}{2.5\times x}$. The resulting \emph{curvilinear polygon} is illustrated on the right with the source of light being the yellow dot.}\label{fig:curves_1}
\end{figure}

Our approach can be extended to generic curves by editing the vector field that produces \eqref{eq:c} while preserving \eqref{eq:CFL}. An example is illustrated in Fig.~\ref{fig:curves_1}. As such, our approach then checks for the existence of \emph{curves-of-sight} rather than lines-of-sight. This can be useful when dealing with a robot that moves in curves rather than straight lines. More complex forms of the underlying vector field can be adopted to possibly reflect robot kinematic constraints. The linear transport equation \eqref{eq:advectionScalarField} can also be generalized to N dimensions \cite{Evans2010}.

Lastly, the approach can check the existence of lines-of-sight and \emph{curves-of-sight} to multiple source points or even a surface $S$. By discretizing the latter, one can evaluate multiple polygons, one for each point, then perform 
\begin{equation}
    \mathcal{U}_{\cap} = \min_{p \in S} \mathcal{U}(p),
    \label{eq:intersection}
\end{equation} 
where $\mathcal{U_{\cap}}$ is the visibility or \emph{curvilinear} polygon common to all points $p$ in $S$, the set of queried points. $\mathcal{U}(p)$ is the polygon per queried point $p$.

\section{Path Planning Using Visibility}
\label{sec:planner}
Visibility through lines-of-sight is the geodesic shortest path between two points and the algorithm introduced in Section ~\ref{sec:visibility} quantifies visibility for the whole grid. It is thus natural to apply Alg.~\ref{alg:VisibilityAlgorithm} to a path planner. 
\subsection{The Visibility Heuristic}
\label{subsec:heuristic}
Let $\mathcal{S}_{0}$ be the set of points having visibility values $\geq 0.5$ in the visibility polygon $\mathcal{U}(p_{0})$ of a starting point $p_{0}$. As such, $\mathcal{S}_{0}$ represents the set of points visible to $p_{0}$. Let $p_{1}$ be a point in $\mathcal{S}_{0}$ having the minimum visibility value and evaluate the visibility polygon $\mathcal{U}(p_{1})$. Performing the union of $\mathcal{U}(p_{0})$ and $\mathcal{U}(p_{1})$ results in the cumulative visibility polygon $\mathcal{U}_{\cup}$ that captures the collective visibility of points $p_{0}$ and $p_{1}$. Repeating the same procedure iteratively allows us to build 
\begin{equation}
    \mathcal{U}_{\cup} = \max_{w \in \mathcal{W}} \mathcal{U}(w),
    \label{eq:union}
\end{equation} 
where $\mathcal{W}$ is the set of waypoints that have been placed so far and $\mathcal{U}_{\cup}$ is the cumulative visibility seen by waypoints in $\mathcal{W}$. Using \eqref{eq:union} allows us to overcome local minima by choosing the next waypoint to be the point in the visible set $\mathcal{S}$ of $\mathcal{U}_{\cup}$ that has minimum visibility value. As such, upcoming waypoints will not be placed in regions that have already been fully explored, but rather in regions that are \emph{barely} visible to $\mathcal{W}$ (meaning points having visibility $v$ such that $\text{threshold} = 0.5 \leq v \ll 1$). The algorithm will keep exploring until the entire map is visible to at least one waypoint. Therefore, the algorithm is guaranteed to establish visibility with all points in the grid that are reachable by straight lines, eventually leading to establishing visibility with the target. 
Assuming that the planner's target point $p_{t}$ does not trivially belong to $\mathcal{S}_{0}$, a terminating criterion is $\mathcal{U}_{\cup_{p_{t}}} \geq 0.5$. This notation indicates that the target point $p_{t}$ has a collective visibility value of at least 0.5 making it visible to at least one waypoint in $\mathcal{W}$. An upper bound on the number of iterations is also set as a terminating criterion for the case where the grid has inaccessible regions.

We add to the scheme a distance bias towards the target in order to pick the next waypoint in a manner that minimizes the path length covered by $\mathcal{W}$ so far, similar to an $A^\ast$ path planner. By saving each waypoint's parent, we also keep track of the path as it is being constructed. The next chosen waypoint $w_{i+1}$ need not have the predecessor waypoint $w_{i}$ as its parent due to the fact that waypoints are chosen based on $\mathcal{U}_{\cup}$ rather than $\mathcal{U}(p_{i})$. Every newly explored point seen by $w_{i}$ but not previously seen by any predecessor gets $w_{i}$ as its parent. This is done in step 5 in Alg.~\ref{alg:PlannerAlgorithm}. 
Step 7 in Alg.~\ref{alg:PlannerAlgorithm} entails picking the point which minimizes the heuristic $\mathcal{H}$
\begin{equation}
    \mathcal{H}_{p} = d_{\text{total}_{\text{p}}} + \Bar{\mathcal{U}}_{\cup_{p}}, \forall p : \mathcal{U}_{\cup_{p}} \geq 0.5,
    \label{eq:heuristic}
\end{equation}
where $d_{\text{total}} = d_{\text{parent}} + d_{\text{target}}$, with $d_{\text{parent}}$ being the distance between a point $p$ and its parent and $d_{\text{target}}$ the distance between $p$ and $p_{t}$. $\Bar{\mathcal{U}}_{\cup_{p}}$ is the properly scaled visibility value at $p$. Our heuristic implementation \eqref{eq:heuristic} may not be the best, but the purpose of this paper is not to produce the best implementation, rather, it is to introduce the proposed visibility \textbf{quantity} as a heuristic for path planning applications.

\begin{algorithm}
\caption{Visibility Heuristic Path Planner}
\label{alg:PlannerAlgorithm}
\textbf{Inputs:} 
$\mathcal{O} \gets$  Probabilistic occupancy grid complement \\
\hspace*{10mm} $p_{0} \gets$ Start position,  $p_{t} \gets$ Target position \\ 
\hspace*{10mm} $(n_{x}, n_{y}) \gets$ Grid dimensions, $\text{max}_{i} \gets$ max iterations \\
\textbf{Output: } $ \mathcal{E} \gets$ Map containing the parents of explored points
\begin{algorithmic}[1]
    \Procedure{VisibilityPathPlanner}{Inputs}
    \State Initialize waypoint $w \gets p_{0}$, $\mathcal{W}$ empty set, $i = 0$
    \While{$\mathcal{U}_{p_{t}} < \text{threshold} \land i < \text{max}_{i}$ }
    \State add $w$ to $\mathcal{W}$
    \State compute $\mathcal{U}(w)$ and update $\mathcal{E}$
    \State $\displaystyle \mathcal{U}_{\cup} \gets \max_{\forall w \in \mathcal{W}} \mathcal{U}(w)$
    \State $\displaystyle w \gets \argmin_{p} \mathcal{H}$ (maintained by a heap)
    \State $i \gets i + 1$
    \EndWhile
    \State \textbf{return} $\mathcal{E}$
    \EndProcedure
\end{algorithmic}
\end{algorithm}

\subsection{Visibility Heuristic Path Planning Results}
\label{subsec:plannerResults}

\begin{figure}[t]
    \centering
    \begin{subfigure}{}
        \includegraphics[width=4cm]{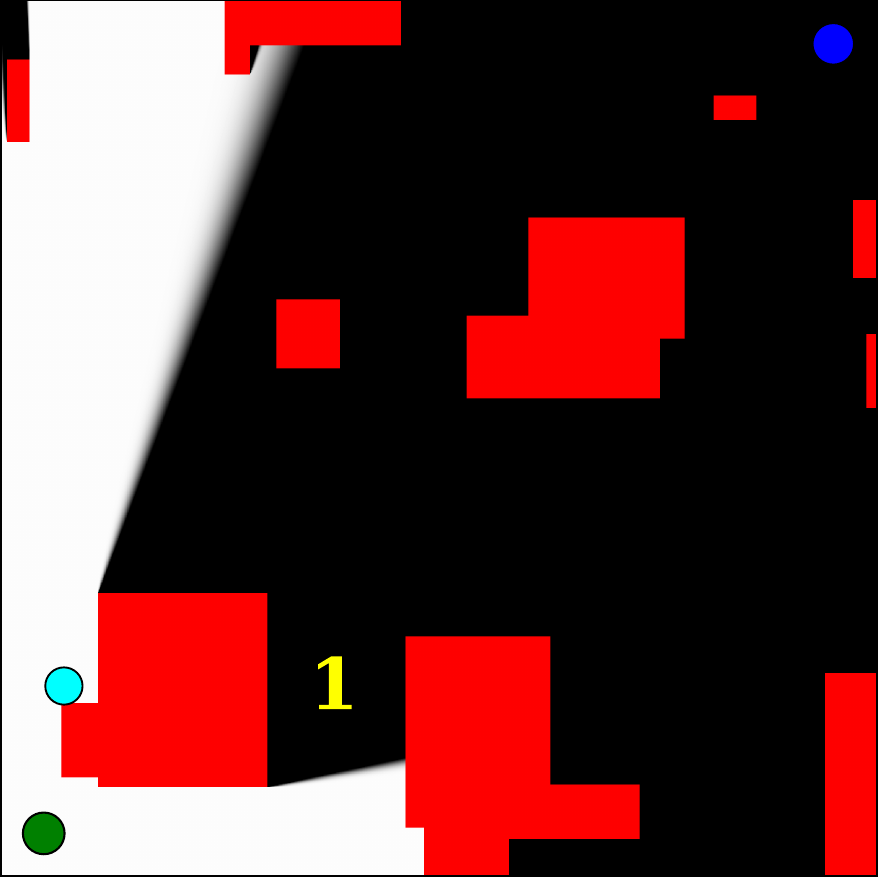}
    \end{subfigure}\hspace*{\fill}
    \begin{subfigure}{}
        \includegraphics[width=4cm]{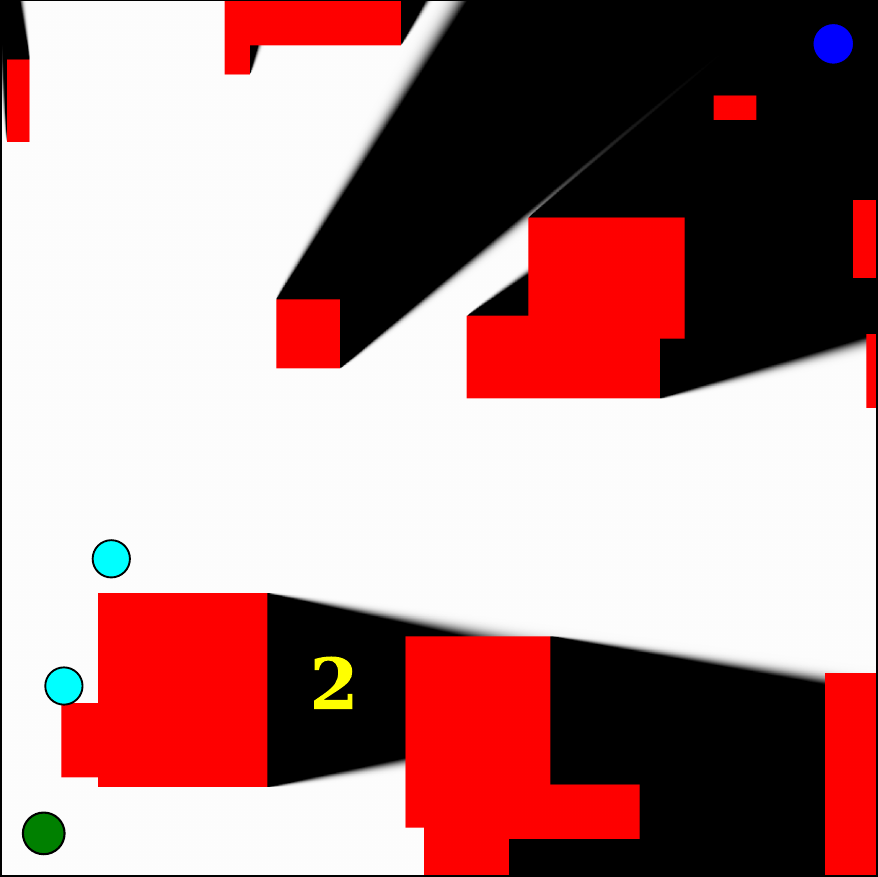}
    \end{subfigure}
    \begin{subfigure}{}
        \includegraphics[width=4cm]{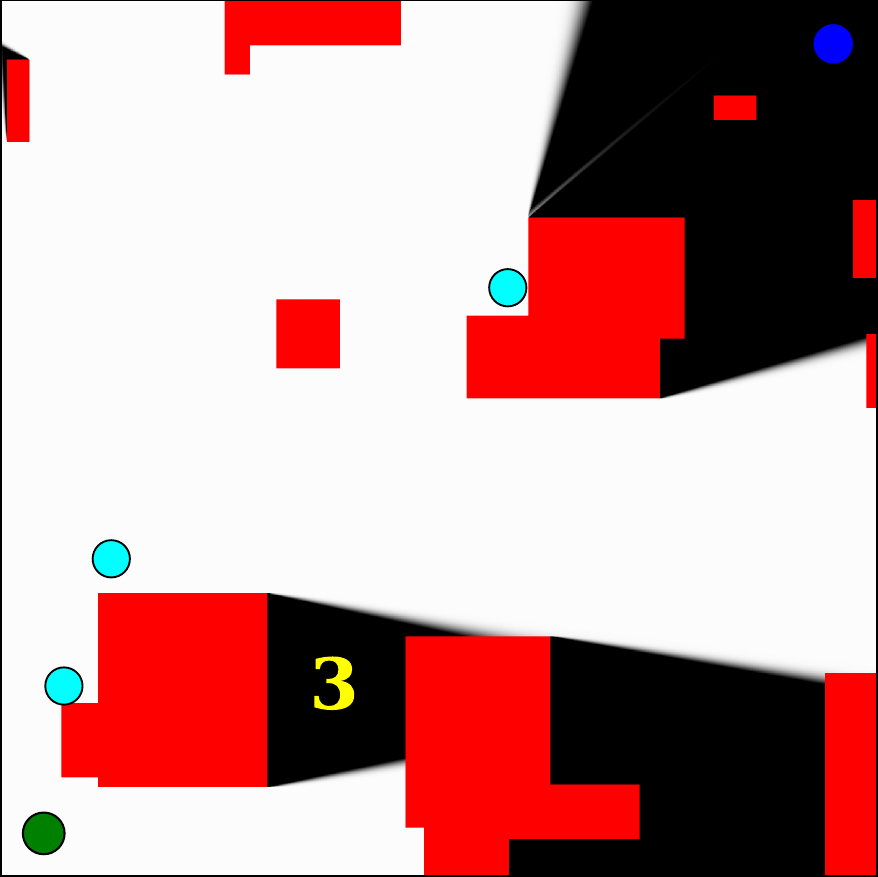}
    \end{subfigure}\hspace*{\fill}
    \begin{subfigure}{}
        \includegraphics[width=4cm]{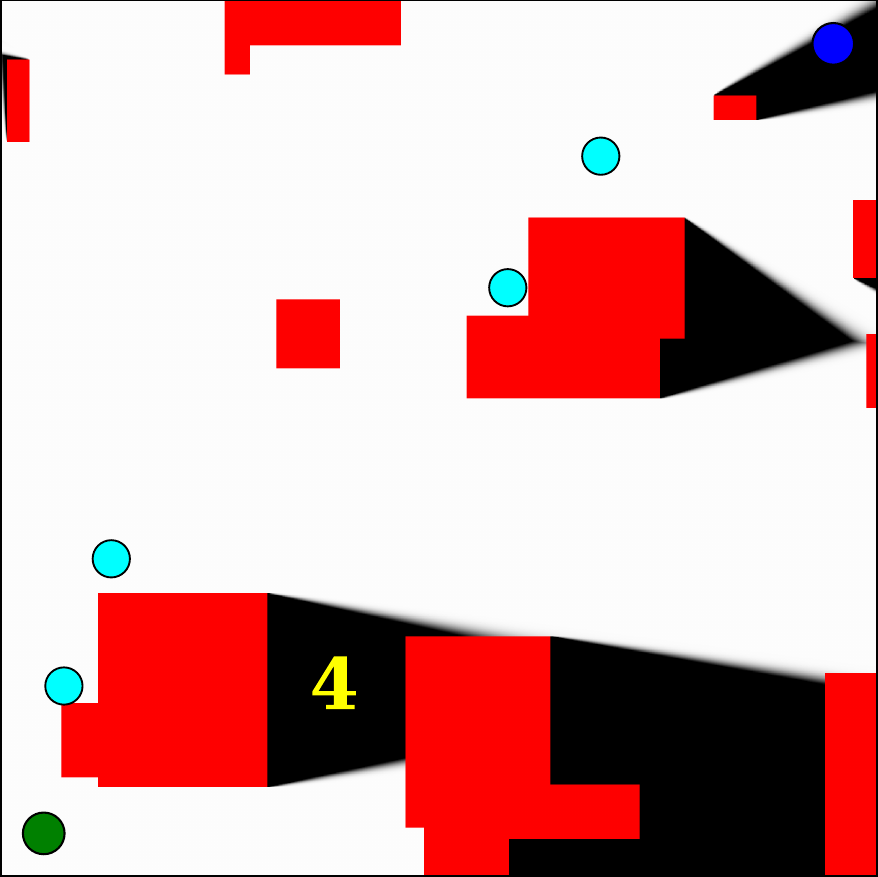}
    \end{subfigure}
    \begin{subfigure}{}
        \includegraphics[width=4cm]{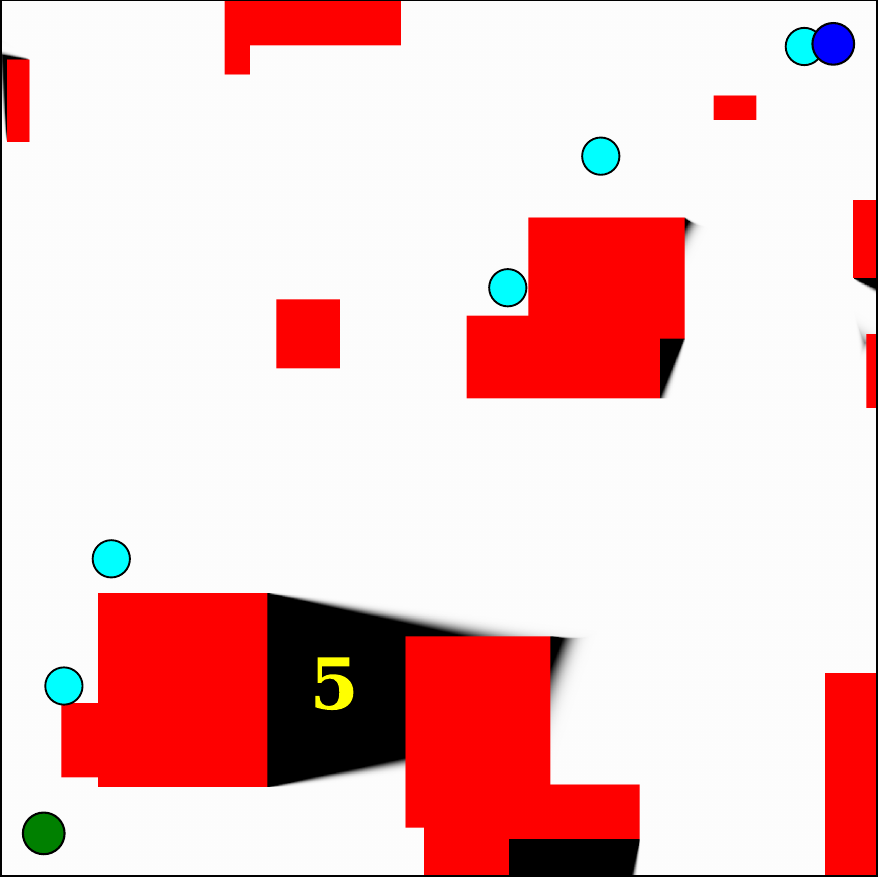}
    \end{subfigure}\hspace*{\fill}
    \begin{subfigure}{}
        \includegraphics[width=4cm]{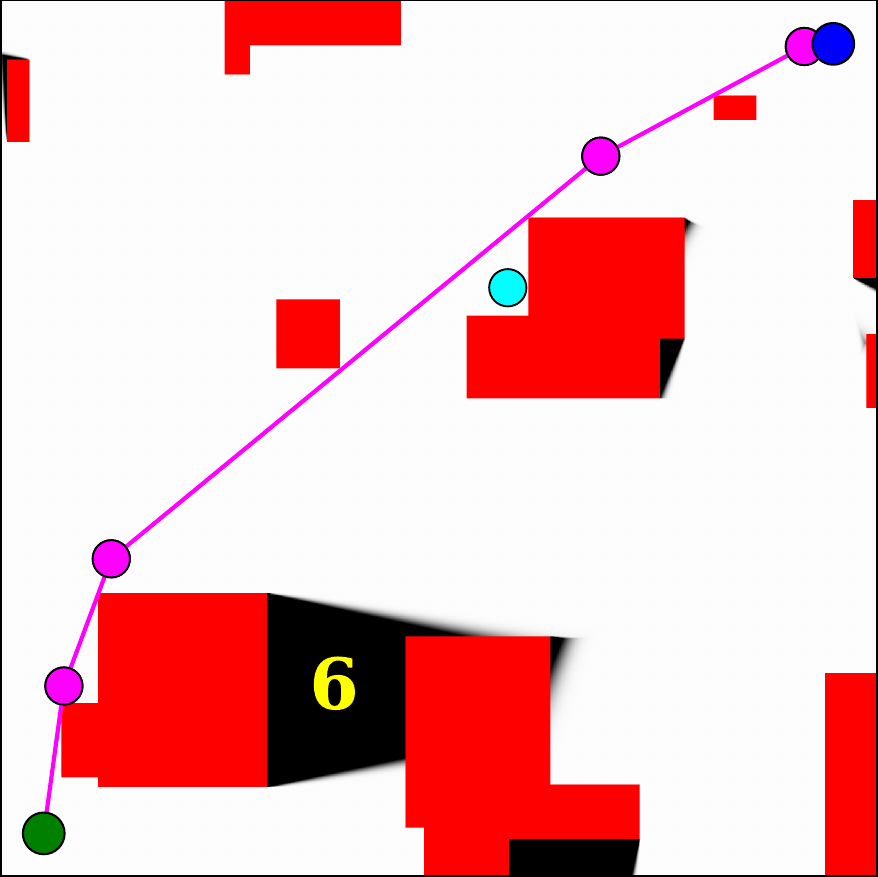}
    \end{subfigure}
    \caption{Progression of the path planner in six steps. Starting point is in green at the bottom left, whereas the target point is the blue one at the top right. Points in cyan are intermediate waypoints, whereas the points and lines in magenta constitute the final path. Obstacles are shown in red.}\label{fig:steps}
\end{figure}

\begin{figure}[ht!]
    \centering
    \begin{subfigure}{}
        \includegraphics[width=4.1cm]{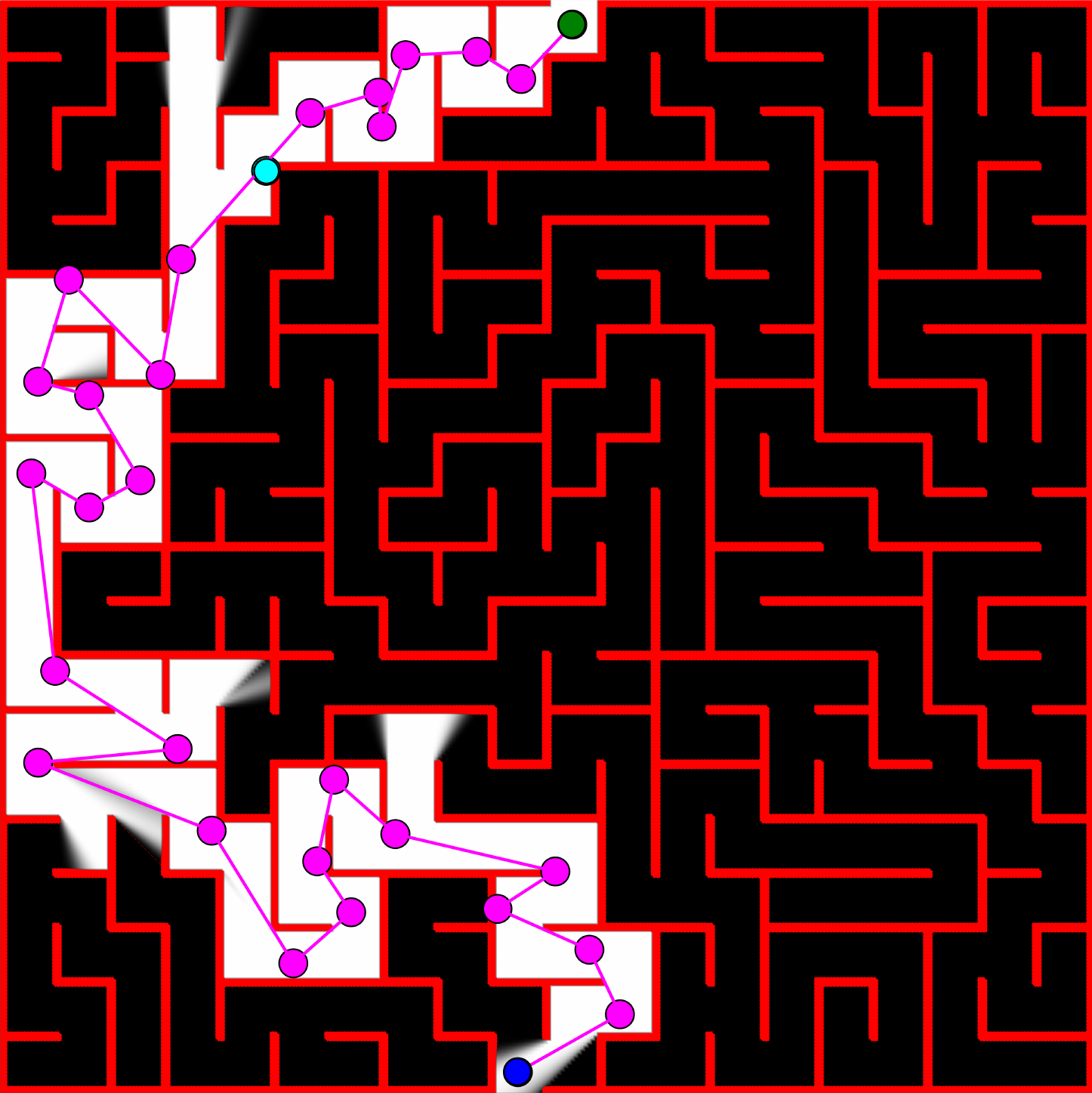}
    \end{subfigure}\hfil
    \begin{subfigure}{}
        \includegraphics[width=4.1cm]{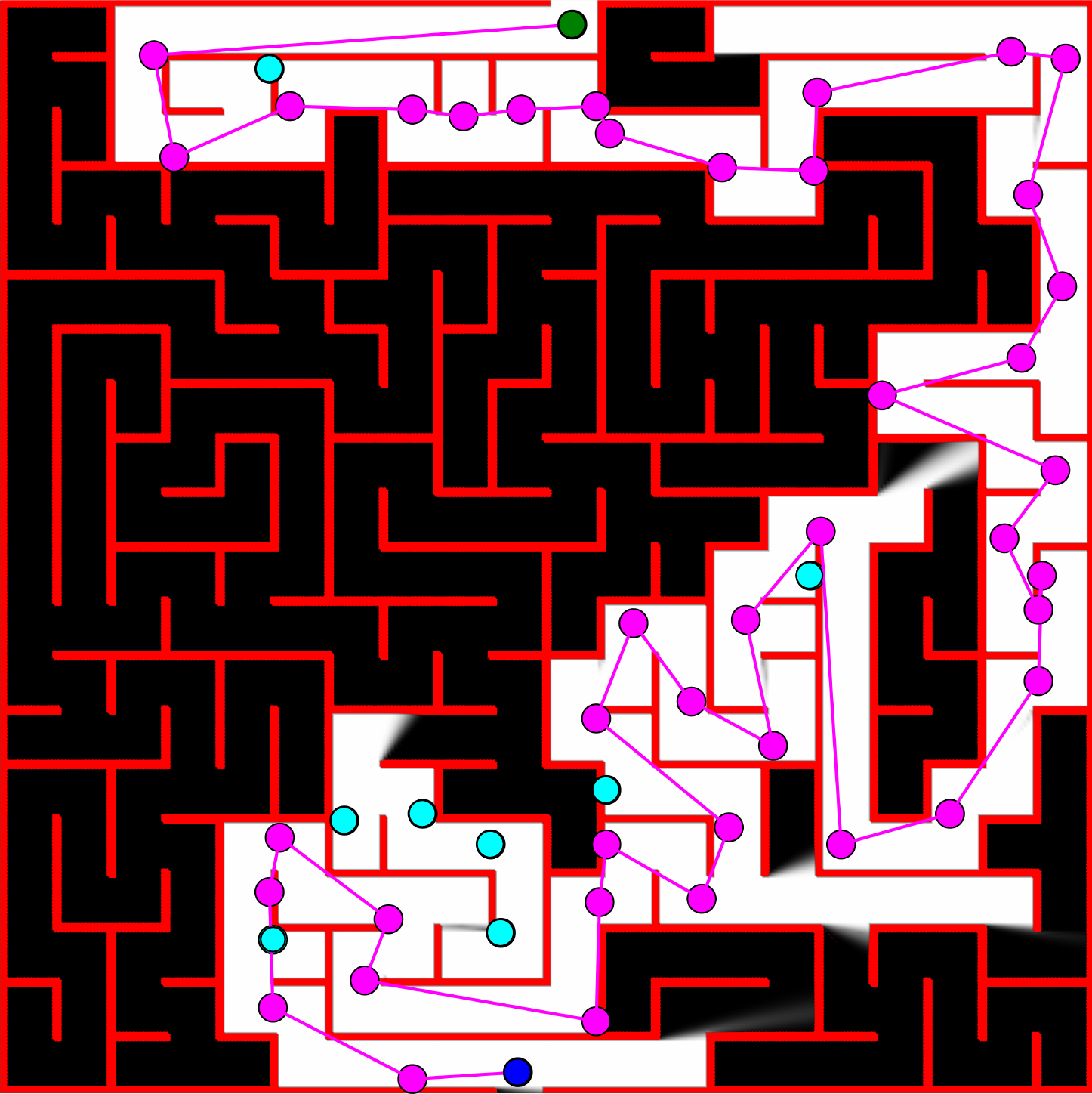}
    \end{subfigure}\hfil
    \caption{Visibility heuristic planner solution in mazes. Start point at top center in green, target point at bottom center in blue, intermediate exploration waypoints in cyan, and final path in magenta. Obstacles in red.}\label{fig:mazes}
\end{figure}

The progression of a solution utilizing Alg.~\ref{alg:PlannerAlgorithm} with a threshold of 0.5 is illustrated in Fig.~\ref{fig:steps} for a $1000\times1000$ grid. In step 1, the visibility polygon for $p_{0}$ is evaluated and the cyan waypoint is chosen based on \eqref{eq:heuristic}. In steps 2-5, the process is repeated while avoiding local minima due to \ref{eq:union}. The stopping criterion is reached by step 6. The resulting path is a point-to-point path. A much more complex maze environment of size $322\times322$ can be explored and a solution can be found quickly with a threshold of 0.2 as illustrated in Fig.~\ref{fig:mazes}. The solution to the left side maze was obtained in \SI{51}{\milli\second} in C++ on an Intel\textregistered~ Core\texttrademark~ i7-9750H Processor. Based on our tests, such an environment is more challenging than a highly-cluttered randomly-generated environment containing curved obstacles.

Alg.~\ref{alg:PlannerAlgorithm}	  scales \emph{linearly} with the number of exploration iterations, which itself depends on the heuristic implementation and tuning choice \eqref{eq:heuristic}. In our implementation, and at every iteration, we are evaluating visibility for the entire grid, even for regions far beyond the visible ones (e.g., dark regions in Fig.~\ref{fig:mazes}). One improvement could be having an inner-loop exist strategy where the algorithm stops evaluating visibility once it goes deep beyond visible regions. The proposed planner is distance sub-optimal when compared to other state-of-the-art any-angle path planners such as Anya \cite{haraborAnya, Uras2015AnEC}.

\section{Conclusion \& Future Work}
\label{sec:conclusion}
In this paper, we introduced a fast and efficient method to evaluate 2D visibility for grid maps using the linear transport equation --- a linear first-order hyperbolic PDE. The proposed method computes both the visibility polygon by evaluating lines-of-sight and the \emph{curvilinear polygon} by evaluating \emph{curves-of-sight}. We demonstrated the efficiency of such an algorithm and we demonstrated its efficacy by introducing an interesting application that uses visibility as a heuristic for path planning. We produced simulation results using the path planner. We also provided pseudo-codes and sample implementations as an open-source code.

As briefly discussed in Section ~\ref{subsec:plannerResults}, our heuristic implementation choice has room for improvement. The number of computations when evaluating visibility at every iteration may be vastly reduced by having stopping criterion reliant on the amount of visibility information being added (using local or on-demand visibility evaluations). A superior path planning algorithm relying on the concept of visibility that we introduced can also be produced. The potential of customizing the shape of the desired curves when using Alg.~\ref{alg:VisibilityAlgorithm} will be explored. Alg.~\ref{alg:VisibilityAlgorithm} can have applications in localization by performing a maximum likelihood estimation based on comparing sensor data, e.g., LIDAR, to computed visibility polygons. Applications in area coverage, sensor placement, and pursuit evasion constitute attractive potential future works relying on Alg.~\ref{alg:VisibilityAlgorithm} or a 3D version of it.

\bibliographystyle{bibtex/IEEEtran} 
\bibliography{bibtex/IEEEabrv, bibtex/bibliography}
 
\end{document}

%% file: misc/mathdef.tex
\newcommand{\argmin}{\mathop{\mathrm{argmin}}}



%% file: bibtex/bibliography.bib
@incollection{Sack2000Domains,
title = {Chapter 6 - Mesh Generation},
editor = {J.-R. Sack and J. Urrutia},
booktitle = {Handbook of Computational Geometry},
publisher = {North-Holland},
address = {Amsterdam},
pages = {291-332},
year = {2000},
isbn = {978-0-444-82537-7},
doi = {https://doi.org/10.1016/B978-044482537-7/50007-3},
url = {https://www.sciencedirect.com/science/article/pii/B9780444825377500073},
author = {Marshall Bern and Paul Plassmann}
}

@article{Bungiu2014EfficientCO,
  title={Efficient Computation of Visibility Polygons},
  author={Francisc Bungiu and Michael Hemmer and John Hershberger and Kan Huang and Alexander Kr{\"o}ller},
  journal={ArXiv},
  year={2014},
  volume={abs/1403.3905}
}

@article{Barbra2013Few,
author = {Barba, Luis and Korman, Matias and Langerman, Stefan and Silveira, Rodrigo I.},
title = {Computing a Visibility Polygon Using Few Variables},
year = {2014},
issue_date = {October, 2014},
publisher = {Elsevier Science Publishers B. V.},
address = {NLD},
volume = {47},
number = {9},
issn = {0925-7721},
doi = {10.1016/j.comgeo.2014.04.001},
journal = {Comput. Geom. Theory Appl.},
month = {oct},
pages = {918–926},
numpages = {9}
}

@article{Chvatal1975Combinatorial,
title = {A combinatorial theorem in plane geometry},
journal = {Journal of Combinatorial Theory, Series B},
volume = {18},
number = {1},
pages = {39-41},
year = {1975},
issn = {0095-8956},
doi = {https://doi.org/10.1016/0095-8956(75)90061-1},
url = {https://www.sciencedirect.com/science/article/pii/0095895675900611},
author = {V Chvátal}
}

@inproceedings{Newell1972Hidden,
author = {Newell, M. E. and Newell, R. G. and Sancha, T. L.},
title = {A Solution to the Hidden Surface Problem},
year = {1972},
isbn = {9781450374910},
publisher = {Association for Computing Machinery},
address = {New York, NY, USA},
url = {https://doi.org/10.1145/800193.569954},
doi = {10.1145/800193.569954},
booktitle = {Proceedings of the ACM Annual Conference - Volume 1},
pages = {443–450},
numpages = {8},
location = {Boston, Massachusetts, USA},
series = {ACM '72}
}

@article{Davis1979ComputationalMO,
  title={Computational Models of Space: Isovists and Isovist Fields},
  author={Larry S. Davis and Michael Benedikt},
  journal={Computer Graphics and Image Processing},
  year={1979},
  volume={11},
  pages={49-72}
}

@article{Joe1987CorrectionsTL,
  title={Corrections to Lee's visibility polygon algorithm},
  author={Barry Joe and R. Bruce Simpson},
  journal={BIT Numerical Mathematics},
  year={1987},
  volume={27},
  pages={458-473}
}

@article{Asano2005VisibilityOD,
  title={Visibility of disjoint polygons},
  author={Takao Asano and Tetsuo Asano and Leonidas J. Guibas and John Hershberger and Hiroshi Imai},
  journal={Algorithmica},
  year={2005},
  volume={1},
  pages={49-63}
}

@article{Heffernan1995Holes,
author = {Heffernan, Paul J. and Mitchell, Joseph S. B.},
title = {An Optimal Algorithm for Computing Visibility in the Plane},
year = {1995},
issue_date = {Feb. 1995},
publisher = {Society for Industrial and Applied Mathematics},
address = {USA},
volume = {24},
number = {1},
issn = {0097-5397},
url = {https://doi.org/10.1137/S0097539791221505},
doi = {10.1137/S0097539791221505},
journal = {SIAM J. Comput.},
month = {feb},
pages = {184–201},
numpages = {18}
}

@article{Bose2002Queries,
title = {Efficient visibility queries in simple polygons},
journal = {Computational Geometry},
volume = {23},
number = {3},
pages = {313-335},
year = {2002},
issn = {0925-7721},
doi = {https://doi.org/10.1016/S0925-7721(01)00070-0},
url = {https://www.sciencedirect.com/science/article/pii/S0925772101000700},
author = {Prosenjit Bose and Anna Lubiw and J.Ian Munro},
}

@Article{Zeng2020View,
author={Zeng, Rui
and Wen, Yuhui
and Zhao, Wang
and Liu, Yong-Jin},
title={View planning in robot active vision: A survey of systems, algorithms, and applications},
journal={Computational Visual Media},
year={2020},
month={Sep},
day={01},
volume={6},
number={3},
pages={225-245},
issn={2096-0662},
doi={10.1007/s41095-020-0179-3},
url={https://doi.org/10.1007/s41095-020-0179-3}
}

@INPROCEEDINGS{Wu2015Active,
author={Wu, Kanzhi and Ranasinghe, Ravindra and Dissanayake, Gamini},
booktitle={2015 IEEE International Conference on Robotics and Automation (ICRA)}, 
title={Active recognition and pose estimation of household objects in clutter}, 
year={2015},
volume={},
number={},
pages={4230-4237},
doi={10.1109/ICRA.2015.7139782}}

@inproceedings{McGreavy2016NBV,
title = "Next Best View Planning for Object Recognition in Mobile Robotics",
author = "Christopher McGreavy and Lars Kunze and Nicholas Hawes",
year = "2016",
month = nov,
day = "10",
language = "English",
booktitle = "Proceedings of the 34th Workshop of the UK Planning and Scheduling Special Interest Group (PlanSIG2016)",
note = "34th Workshop of the UK Planning and Scheduling Special Interest Group (PlanSIG2016) ; Conference date: 15-12-2016 Through 16-12-2016",
}

@Article{Santos2020Autonomous,
AUTHOR = {Santos, João and Oliveira, Miguel and Arrais, Rafael and Veiga, Germano},
TITLE = {Autonomous Scene Exploration for Robotics: A Conditional Random View-Sampling and Evaluation Using a Voxel-Sorting Mechanism for Efficient Ray Casting},
JOURNAL = {Sensors},
VOLUME = {20},
YEAR = {2020},
NUMBER = {15},
ARTICLE-NUMBER = {4331},
ISSN = {1424-8220},
DOI = {10.3390/s20154331}
}

@inproceedings{Bowen2020Emergent,
  author    = {Bowen Baker and
               Ingmar Kanitscheider and
               Todor M. Markov and
               Yi Wu and
               Glenn Powell and
               Bob McGrew and
               Igor Mordatch},
  title     = {Emergent Tool Use From Multi-Agent Autocurricula},
  booktitle = {8th International Conference on Learning Representations, {ICLR} 2020,
               Addis Ababa, Ethiopia, April 26-30, 2020},
  publisher = {OpenReview.net},
  year      = {2020},
  bibsource = {dblp computer science bibliography, https://dblp.org}
}

@inproceedings{Tandon2018MedusaTS,
  title={Medusa: Towards Simulating a Multi-Agent Hide-and-Seek Game},
  author={Akshata Tandon and K. Karlapalem},
  booktitle={International Joint Conference on Artificial Intelligence},
  year={2018}
}

@article{Amanatides1987Traversal,
author = {Amanatides, John and Woo, Andrew},
year = {1987},
month = {08},
pages = {},
title = {A Fast Voxel Traversal Algorithm for Ray Tracing},
volume = {87},
journal = {Proceedings of EuroGraphics}
}

@INPROCEEDINGS{Ibrahim2022Occlusion,
  author={Ibrahim, Ibrahim and Farshidian, Farbod and Preisig, Jan and Franklin, Perry and Rocco, Paolo and Hutter, Marco},
  booktitle={2022 International Conference on Robotics and Automation (ICRA)}, 
  title={Whole-Body MPC and Dynamic Occlusion Avoidance: A Maximum Likelihood Visibility Approach}, 
  year={2022},
  volume={},
  number={},
  pages={221-227},
  doi={10.1109/ICRA46639.2022.9811536}}

@ARTICLE{Nageli2017Aerial,
  author={Nägeli, Tobias and Alonso-Mora, Javier and Domahidi, Alexander and Rus, Daniela and Hilliges, Otmar},
  journal={IEEE Robotics and Automation Letters}, 
  title={Real-Time Motion Planning for Aerial Videography With Dynamic Obstacle Avoidance and Viewpoint Optimization}, 
  year={2017},
  volume={2},
  number={3},
  pages={1696-1703},
  doi={10.1109/LRA.2017.2665693}}

@article{Lozano1979Polyhedral,
author = {Lozano-P\'{e}rez, Tom\'{a}s and Wesley, Michael A.},
title = {An Algorithm for Planning Collision-Free Paths among Polyhedral Obstacles},
year = {1979},
issue_date = {Oct. 1979},
publisher = {Association for Computing Machinery},
address = {New York, NY, USA},
volume = {22},
number = {10},
issn = {0001-0782},
url = {https://doi.org/10.1145/359156.359164},
doi = {10.1145/359156.359164},
journal = {Commun. ACM},
month = {oct},
pages = {560–570},
numpages = {11}
}

@article{Pocchiola1996TopologicallySV,
  title={Topologically sweeping visibility complexes via pseudotriangulations},
  author={Michel Pocchiola and Gert Vegter},
  journal={Discrete \& Computational Geometry},
  year={1996},
  volume={16},
  pages={419-453}
}

@article{Kapoor2000VGraph,
author = {Kapoor, Sanjiv and Maheshwari, S. N.},
title = {Efficiently Constructing the Visibility Graph of a Simple Polygon with Obstacles},
journal = {SIAM Journal on Computing},
volume = {30},
number = {3},
pages = {847-871},
year = {2000},
doi = {10.1137/S0097539795253591},
URL = { 
    
        https://doi.org/10.1137/S0097539795253591
},
eprint = { 
    
        https://doi.org/10.1137/S0097539795253591
}
,
}

@unpublished{Allaire2022Accessibility,
  TITLE = {{Accessibility constraints in structural optimization via distance functions}},
  AUTHOR = {Allaire, Gr{\'e}goire and Bihr, Martin and Bogosel, Beniamin and Godoy, Matias},
  URL = {https://hal.science/hal-03864841},
  NOTE = {working paper or preprint},
  YEAR = {2022},
  MONTH = Nov,
  HAL_ID = {hal-03864841},
  HAL_VERSION = {v1},
}

@book{Evans2010,
title={Partial Differential Equations},
author={Evans, Lawrence C.},
year={2010},
publisher={American Mathematical Society}
}

@article{Lewy1928,
author = {Lewy, Hans and Friedrichs, Kurt and Courant, Rrichard},
journal = {Mathematische Annalen},
pages = {32-74},
title = {Über die partiellen Differenzengleichungen der mathematischen Physik},
url = {http://eudml.org/doc/159283},
volume = {100},
year = {1928},
}

@book{hirsch2007numerical,
  author = {Hirsch, Charles},
  publisher = {Elsevier},
  title = {Numerical computation of internal and external flows: The fundamentals of computational fluid dynamics},
  year = 2007
}

@book{patankar1980numerical,
  author = {Patankar, Suhas V},
  isbn = {978-0891165224},
  publisher = {Hemisphere Publishing Corporation (CRC Press, Taylor \& Francis Group)},
  series = {Series on Computational Methods in Mechanics and Thermal Science},
  title = {Numerical heat transfer and fluid flow},
  url = {http://www.crcpress.com/product/isbn/9780891165224},
  year = 1980
}

@ARTICLE{OPM2020,
  author={Farias, Renato and Kallmann, Marcelo},
  journal={IEEE Transactions on Visualization and Computer Graphics}, 
  title={Optimal Path Maps on the GPU}, 
  year={2020},
  volume={26},
  number={9},
  pages={2863-2874},
  doi={10.1109/TVCG.2019.2904271}
}

@inproceedings{Uras2015AnEC,
  title={An Empirical Comparison of Any-Angle Path-Planning Algorithms},
  author={Tansel Uras and Sven Koenig},
  booktitle={Symposium on Combinatorial Search},
  year={2015}
}

@inproceedings{haraborANYA,
author = {Harabor, Daniel and Grastien, Alban},
year = {2013},
month = {06},
pages = {},
title = {An Optimal Any-Angle Pathfinding Algorithm},
booktitle = {ICAPS 2013 - Proceedings of the 23rd International Conference on Automated Planning and Scheduling}
}

@article{LCT2014,
author = {Kallmann, Marcelo},
title = {Dynamic and Robust Local Clearance Triangulations},
year = {2014},
issue_date = {August 2014},
publisher = {Association for Computing Machinery},
address = {New York, NY, USA},
volume = {33},
number = {5},
issn = {0730-0301},
url = {https://doi.org/10.1145/2580947},
doi = {10.1145/2580947},
journal = {ACM Trans. Graph.},
month = {sep},
articleno = {161},
numpages = {17}
}
